%% file: Paper-v9.tex
\documentclass[
aps,
pre,
reprint,
tightenlines,
superscriptaddress,
amsmath,
amssymb,
noshowkeys,
nofootinbib,
floatfix,
]{revtex4-2}
\input{JohnsREVTEXdefs}

\newcommand{\barPsi}{\bar{\Psi}}
\newcommand{\barPhi}{\bar{\Phi}}

\newcommand{\bq}{\begin{equation}}
\newcommand{\ba}{\begin{eqnarray}}
\newcommand{\eq}{\end{equation}}
\newcommand{\ea}{\end{eqnarray}}
\newcommand{\be}{\begin{equation}}
\newcommand{\ee}{\end{equation}}
\newcommand{\bea}{\begin{eqnarray}}
\newcommand{\eea}{\end{eqnarray}}

\begin{document}
%
%
\preprint{LA-UR-24-26709}
\begin{flushright}\textbf{LA-UR-24-26709}\end{flushright}
\title{Solitary waves in the coupled nonlinear massive Thirring as well as 
coupled Soler models with arbitrary nonlinearity}

\author{Avinash Khare}
\email{avinashkhare45@gmail.com} 
\affiliation{
   Physics Department, 
   Savitribai Phule Pune University, 
   Pune 411007, India} 
\author{Fred Cooper}
\email{cooper@santafe.edu}
\affiliation{
   Santa Fe Institute,
   1399 Hyde Park Road,
   Santa Fe, NM 87501, USA}
\affiliation{
   Center for Nonlinear Studies and Theoretical Division, 
   Los Alamos National Laboratory, 
   Los Alamos, NM 87545, USA}
\author{John F. Dawson}
\email{john.dawson@unh.edu}
\affiliation{
   Department of Physics,
   University of New Hampshire,
   Durham, NH 03824, USA}
\author{Efstathios G. Charalampidis}
\affiliation{Mathematics Department,
California Polytechnic State University,
San Luis Obispo, CA 93407-0403, USA}
\affiliation{Nonlinear Dynamical Systems Group,
Computational Sciences Research Center, and
Department of Mathematics and Statistics,
San Diego State University,
San Diego, CA 92182-7720, USA}
\author{Avadh Saxena} 
\email{avadh@lanl.gov} 
\affiliation{
   Center for Nonlinear Studies and Theoretical Division, 
   Los Alamos National Laboratory, 
   Los Alamos, NM 87545, USA}
\date{\today, \now \ PST}
\begin{abstract}
Motivated by the recent introduction of an integrable coupled massive Thirring 
model by Basu-Mallick et al, we introduce a new coupled Soler model. Further we 
generalize both the coupled massive Thirring and the coupled Soler model to 
arbitrary nonlinear parameter $\kappa$ and obtain exact solitary wave solutions
in both cases. Remarkably, it turns out that in both the models, because of
the conservation laws of charge and energy, the exact solutions we find seem to
not depend on how we parameterize them, and the charge density of these solutions
is related to the charge density of the single field solutions found earlier
by a subset of the present authors. In both the models, a nonrelativistic reduction
of the equations leads to the same conclusion that the solutions are proportional
to those found in the one component field case.
\end{abstract}
\maketitle
%
%
%
%
\section{\label{s:Intro}Introduction}

Recently a new integrable coupled massive Thirring model (MTM) \cite{Thirring-1958} in $1+1$ dimensions with field variables in a complex algebra has been introduced by Basu-Mallick, et~al.~\cite{BasuMallick2023bosonic}.  It is then natural to inquire if similarly one can also introduce a coupled Soler model \cite{Soler-1970}. In this context it is worth recalling that few years ago, we \cite{PhysRevE.82.036604} generalized the uncoupled (single component) MTM as well as Soler model to arbitrary nonlinearity $\kappa$ and obtained the solitary wave solutions in both cases.

The purpose of this paper is to introduce a coupled Soler model and further extend both the coupled MTM and the coupled Soler model to arbitrary nonlinearity $\kappa$, and try to obtain solitary wave solutions of both coupled models. It is worth recalling here that whereas in MTM one has vector-vector (v-v) nonlinear coupling, in the Soler model one has scalar-scalar (s-s) nonlinear coupling.

The plan of the paper is as follows. In Section~\ref{s:NLDE} we introduce the Lagrangian for the two-component generalized nonlinear Dirac (NLD) model for both the, and derive equations of motion and conservation equations for the model. We find that the solitary wave solutions in the rest system are constrained by the laws of
momentum and energy so that the solutions are proportional to what was found in the one-component case. Therefore we expect that the stability property of these solutions may not be very different than those we found for the solitary waves in the one field case.  The moving solitary waves can be generated from the solutions we found by the appropriate Lorentz boost. 

%
%
\section{\label{s:NLDE}Coupled Dirac equation with s-s and v-v couplings}

In this section we introduce the coupled Soler model which has s-s nonlinear
coupling and consider both the coupled models together since the only 
difference between them is the nonlinear s-s or v-v coupling. We extend both the
models to arbitrary nonlinearity $\kappa$ and derive general properties in 
the case of both the models.

The Lagrangian for both the models is given by
\begin{equation}\label{e:Lagrangian}
   L
   =
   \frac{1}{2} \, 
   \bigl \{ \barPhi (\rmi \gamma^{\mu} \partial_{\mu} - m) \Psi 
   +
   \barPsi (\rmi \gamma^{\mu} \partial_{\mu} - m) \Phi
   +
   \text{h.c.} \bigr \}
   +
   L_I \>,
\end{equation}
where
\begin{equation}\label{e:LI-ss}
   L_I
   =
   \frac{g^2}{\kappa+1}
   \bigl \{\, (\barPhi \Psi)^{\kappa+1} + (\barPsi \Phi)^{\kappa+1} \, \bigr \} \>,
\end{equation}
for the s-s case, and
\begin{align}\label{e:LI-vv}
   L_I
   &=
   \frac{g^2}{\kappa+1}
   \bigl \{\, 
      [ (\barPhi \gamma_{\mu} \Psi) (\barPhi \gamma^{\mu} \Psi) ]^{(\kappa+1)/2}
      \\
      & \hspace{3em}
      + 
      [ (\barPsi \gamma_{\mu} \Phi) (\barPsi \gamma^{\mu} \Phi) ]^{(\kappa+1)/2} \,
   \bigr \}
   \notag
\end{align}
for the v-v case.
The Lagrangian for both the models is invariant under local Lorentz 
transformations.

For s-s coupling, the equations of motion are given by
\begin{subequations}\label{e:ss-eom}
\begin{align}
   (\rmi \gamma^{\mu} \partial_{\mu} - m) \Psi 
   + 
   g^2 (\barPhi \Psi)^{\kappa} \, \Psi
   &= 
   0 \>,
   \label{e:ss-eom-a} \\
   (\rmi \gamma^{\mu} \partial_{\mu} - m) \Phi 
   + 
   g^2 (\barPsi \Phi)^{\kappa} \, \Phi
   &= 
   0 \>,
   \label{e:ss-eom-b}   
\end{align}
\end{subequations}
whereas for v-v coupling, the equations of motion are
\begin{subequations}\label{e:vv-eom}
\begin{align}
   &(\rmi \gamma^{\mu} \partial_{\mu} - m) \Psi 
   \label{e:vv-eom-a} \\
   & \hspace{1em}
   + 
   g^2 [(\barPhi \gamma_{\nu} \Psi)
   (\barPhi \gamma^{\nu} \Psi) ]^{(\kappa-1)/2}
   (\barPhi \gamma_{\mu} \Psi)
   \gamma^{\mu} \Psi
   = 
   0 \>,
   \notag \\
   &(\rmi \gamma^{\mu} \partial_{\mu} - m) \Phi 
   \label{e:vv-eom-b}\\
   & \hspace{1em}
   + 
   g^2 [(\barPsi \gamma_{\nu} \Phi)
   (\barPsi \gamma^{\nu} \Phi) ]^{(\kappa-1)/2}
   (\barPsi \gamma_{\mu} \Phi)
   \gamma^{\mu} \Phi
   = 
   0 \>,
   \notag   
\end{align}
\end{subequations}
For both the s-s and the v-v couplings, current is conserved:
\begin{equation}\label{e:currentcons}
   \partial_{\mu} j^{\mu}(x) = 0
   \qc
   j^{\mu} = \frac{1}{2} \, \{\, \barPhi \gamma^{\mu} \Psi + \barPsi \gamma^{\mu} \Phi\, \} \>.
\end{equation}
This means that the charge $Q$ is independent of time, where
\begin{equation}\label{e:Qdef}
   Q
   =
   \int \dd[3]{x} j^{0}(x)
   =
   \frac{1}{2} \int \dd[3]{x} 
   \{\, \Phi^{\dag} \Psi + \Psi^{\dag} \Phi \, \}\>.
\end{equation}
The stress-energy tensor is also conserved:
\begin{equation}\label{e:SEtensorCons}
   \partial_{\mu} T^{\mu\nu}(x) = 0 \>,
\end{equation}
where
\begin{gather}\label{e:Tdef}
   T^{\mu\nu} = \frac{1}{2} \, \{\, D^{\mu\nu} + \text{h.c.} \,\} - g^{\mu\nu} L \>,
   \\
   D^{\mu\nu} 
   =
   \rmi \, \barPhi \gamma^{\mu} \partial^{\nu} \Psi
   +
   \rmi \, \barPsi \gamma^{\mu} \partial^{\nu} \Phi \>,
\end{gather}
which means that the linear momentum vector $P^{\nu} = (E,\vb{P})$ is conserved:
\begin{equation}\label{e:Jmu-def}
   P^{\nu}
   =
   \int \dd[3]{x} T^{0 \nu}(x)\>.
\end{equation}
We will use these results in the following sections.

%
%
\subsection{\label{ss:TwoDimensions}Reduction to $1+1$ dimensions}

In two dimensions with $\dd{x^{\mu}} = ( \dd{t},\dd{x})$, we use the representations given in Ref.~\onlinecite{PhysRevE.82.036604} where
\begin{equation}\label{e:gammaReps}
   \gamma^{0} = \sigma_3 = 
   \begin{pmatrix} 1 & 0 \\ 0 & -1 \end{pmatrix}
   \qc
   \gamma^{1} = \rmi \sigma_1 = 
   \begin{pmatrix} 0 & \rmi \\ \rmi & 0 \end{pmatrix} \>,
\end{equation}
and where $\sigma_i$ are the Pauli matrices.  The gamma matrices then obey the anti-commutation relation, $\{ \gamma^{\mu}, \, \gamma^{\nu} \} = 2 g^{\mu\nu}$.  

In both the s-s and v-v model versions we look for solutions in the solitary wave rest frame of the form,
\begin{equation}\label{e:timedepend}
   \Psi(x,t) = \begin{pmatrix} u_1(x) \\ u_2(x) \end{pmatrix} \, \rme^{-\rmi \omega t}
   \qc
   \Phi(x,t) = \begin{pmatrix} v_1(x) \\ v_2(x) \end{pmatrix} \, \rme^{-\rmi \omega t} \>,
\end{equation}
with $u_i(x)$ and $v_i(x)$ real functions with the properties that $u_i(x) \rightarrow 0$ and $v_i(x) \rightarrow 0$ as $|x| \rightarrow \infty$.  This makes the equations of motion real, but the coordinate components of the current density and stress-energy tensor are imaginary.  However we shall see that the boundary conditions of the solitons at infinity will require these quantities to vanish identically.  Moving solitary waves are obtained from this solution by a Lorentz boost.
The s-s equations of motion \eqref{e:ss-eom} becomes:
\begin{subequations}\label{e:ss-eom-II}
\begin{align}
   u'_1 + (m + \omega) u_2 - g^2 (v_1 u_1 - v_2 u_2 )^{\kappa} \, u_2
   &=
   0 \>,
   \label{e:ss-eom-II-a} \\
   u'_2 + (m - \omega) u_1 - g^2 (v_1 u_1 - v_2 u_2 )^{\kappa} \, u_1
   &=
   0 \>,
   \label{e:ss-eom-II-b} \\ 
   v'_1 + (m + \omega) v_2 - g^2 (v_1 u_1 - v_2 u_2 )^{\kappa} \, v_2
   &=
   0 \>,
   \label{e:ss-eom-II-c} \\
   v'_2 + (m - \omega) v_1 - g^2 (v_1 u_1 - v_2 u_2 )^{\kappa} \, v_1
   &=
   0 \>,
   \label{e:ss-eom-II-d} 
\end{align}
\end{subequations}

For the v-v case, the  equations of motion are:
\begin{subequations}\label{e:vv-eom-II}
\begin{align}
   &u'_1 + (m + \omega) u_2  \nonumber \\
   & + 
   g^2 (u_1^2 + u_2^2 )^{(\kappa+1)/2} (v_1^2 + v_2^2 )^{(\kappa-1)/2} \, v_2
   =
   0 \>,
   \label{e:vv-eom-II-a} \\
   & u'_2 + (m - \omega) u_1  \nonumber \\
   & - 
   g^2 (u_1^2 + u_2^2 )^{(\kappa+1)/2} (v_1^2 + v_2^2 )^{(\kappa-1)/2} \, v_1
   =
   0 \>,
   \label{e:vv-eom-II-b} \\ 
   & v'_1 + (m + \omega) v_2  \nonumber \\
   &+ 
   g^2 (v_1^2 + v_2^2 )^{(\kappa+1)/2} (u_1^2 + u_2^2 )^{(\kappa-1)/2} \, u_2
   =
   0 \>,
   \label{e:vv-eom-II-c} \\
   & v'_2 + (m - \omega) v_1  \nonumber \\
   & -
   g^2 (v_1^2 + v_2^2 )^{(\kappa+1)/2} (u_1^2 + u_2^2 )^{(\kappa-1)/2} \, u_1
   =
   0 \>.
   \label{e:vv-eom-II-d} 
\end{align}
\end{subequations}

The zero component of the current (the pair density) is real is given by
\begin{align}\label{e:rhox}
   \rho(x) 
   &=
   \frac{1}{2} \{\, \Phi^{\dag}(x) \Psi(x) + \Psi^{\dag}(x) \Phi(x) \, \}
   \\
   &= 
   v_1(x) u_1(x) + v_2(x) u_2(x) \>,
   \notag
\end{align}
and the charge by the integral: $Q = \int \dd{x} \rho(x)$. The space component of the current is
\begin{equation}\label{e:j1}
   j^{1}(x)
   =
   \rmi \, \{\, u_1(x) \,v_2(x) - u_2(x) \,v_1(x) \} \>.
   \notag
\end{equation}
Current conservation then becomes:
\begin{equation}\label{e:current-III}
   \partial_{x} [\, u_1(x) \,v_2(x) - u_2(x) \,v_1(x) \,] = 0 \>.
\end{equation}
But since we require that $u_i(x) \rightarrow 0$ and $v_i(x) \rightarrow 0$ as $|x| \rightarrow \pm \infty$, the current vanishes identically and we have that
\begin{equation}\label{e:condition}
   u_1(x) \, v_2(x) = u_2(x)\, v_1(x) \>.
\end{equation}
This is a very severe constraint on the solution since it implies
\begin{equation} 
\frac{u_2(x)}{u_1(x)} = \frac{v_2(x)}{v_1(x)}= g(x) \>.
\end{equation} 
It will be convenient to choose this ratio to define a new variable $\theta(x)$ by
\begin{equation} 
g(x) = \tan[\theta(x)],\>.
\end{equation}

The zero component of the linear momentum vector is conserved and the space components vanish identically.  We derive these components in Appendix~\ref{s:StressEnergyTensor}, where from \eqref{e:T11-ss} we find for the s-s model, 
\begin{align}
   &
   \omega \, (u_1 v_1 + u_2 v_2)
   -
   m \, (u_1 v_1 - u_2 v_2)
   \label{e:T11ss-conserved} \\
   & \hspace{3em}
   +
   \frac{g^2}{\kappa+1}\, (v_1 u_1 - v_2 u_2)^{\kappa+1}
   =
   0 \>,
   \notag
\end{align}
whereas from \eqref{e:T11-vv} for the v-v model, we find
\begin{align}
   &\omega \, (u_1 v_1 + u_2 v_2)
   -
   m \, (u_1 v_1 - u_2 v_2)
   \label{e:T11vv-conserved} \\
   & \hspace{1em}
   +
   \frac{g^2}{\kappa+1}
   [ (u_1^2 + u_2^2)(v_1^2 + v_2^2) ]^{(\kappa+1)/2}
   =
   0 \>.
   \notag
\end{align}
We use these results in the following sections.

%
%
\subsection{\label{ss.RadialForm}Radial form}

Without loss of generality we can introduce the variables $R_u$, $R_v$ as
\begin{align}
   u_1^2 + u_2^2 &= R_u^2 \>,
   &
   v_1^2 + v_2^2 &= R_v^2 
\end{align}
and 
\begin{align}
 \tan(\theta_u) &= u_2/u_1 \>,
   &
   \tan(\theta_v) &= v_2/v_1 \>.
   \notag
\end{align}
We have required $u_i(x)$ and $v_i(x)$ to be real functions, so let us set
\begin{align}\label{e.1:RadialForm}
   u_1 &= R_{u} \cos(\theta_u) \>,
   &
   u_2 &= R_{u} \sin(\theta_u) \>,
   \\
   v_1 &= R_{v} \cos(\theta_v) \>,
   &
   v_2 &= R_{v} \sin(\theta_v) \>.
   \notag
\end{align}
Because of the constraint Eq. \eqref{e:condition}, we must have
\begin{equation}
  \tan(\theta_u) = \tan(\theta_v) = \tan(\theta)\>.
\end{equation}
We find that two of the quantities that appear in $T_{11}$ can be written as:
\begin{align}
   u_1 v_1 + u_2 v_2 &= R_u R_v \>,
   \\
   u_1 v_1 - u_2 v_2 &= R_u R_v \cos(2 \theta) \>.
   \notag
\end{align}
Derivatives are now given by:
\begin{subequations}\label{e.1:partials}
\begin{align}
   u'_1
   &=
   R'_{u} \cos(\theta) - R_{u} \theta' \sin(\theta) \>,
   \label{e.1:partials-a} \\
   u'_2
   &=
   R'_{u} \sin(\theta) + R_{u} \theta' \cos(\theta) \>,
   \label{e.1:partials-b} \\
   v'_1
   &=
   R'_{v} \cos(\theta) - R_{v} \theta' \sin(\theta) \>,
   \label{e.1:partials-c} \\
   v'_2
   &=
   R'_{v} \sin(\theta) + R_{v} \theta' \cos(\theta) \>.
   \label{e.1:partials-d}   
\end{align}
\end{subequations}
We then find that
\begin{subequations}\label{e:dervidents}
\begin{align}
   u'_1 v_2 - u'_2 v_1 &= - R_u R_v \theta' \>,
   \\
   v'_1 u_2 - v'_2 u_1 &= - R_u R_v \theta' \>.
\end{align}
\end{subequations}
The charge density \eqref{e:rhox} is then given by
\begin{equation}\label{e:rhoRR}
   \rho(x) = R_u(x) R_v(x) \>,
\end{equation}
and is independent of the phase angle.  
From \eqref{e:T11-ss} for s-s coupling, the $T^{11}$ components of the stress-energy tensor requires that
\begin{equation}\label{e:T11ss-II}
   \omega
   -
   m \, \cos(2 \theta)
   +
   \frac{g^2}{\kappa+1}\, [ R_u R_v ]^{\kappa}
   \cos^{\kappa+1}(2 \theta)
   =
   0 \>,
\end{equation}
and from \eqref{e:T11-vv} for v-v coupling, requires that
\begin{equation}\label{e:T11vv-II}
   \omega 
   -
   m \, \cos(2 \theta)
   +
   \frac{g^2}{\kappa+1} \, [R_u R_v]^{\kappa}
   =
   0 \>.
\end{equation}
From \eqref{e:T00-ss}, the energy density for s-s coupling is given by
\begin{align}
   \varepsilon_{\text{s-s}}(x)
   &\equiv
   T^{00}_{\text{s-s}}(x)
   \label{e:T00ss-II} \\
   & \hspace{-3em}
   =
   2 R_u R_v \, 
   \Bigl \{\, 
      m \cos(2 \theta)
      + 
      \theta' 
      - 
      \frac{g^2 [ R_u R_v\cos(2 \theta) ]^{\kappa+1}}{\kappa+1}\, 
       \,
   \Bigr \} \>,
   \notag
\end{align}
and from \eqref{e:T00-vv} for the v-v model, the energy density is
\begin{align}
   \varepsilon_{\text{v-v}}(x)
   &\equiv
   T^{00}_{\text{v-v}}(x)
   \label{e:T00vv-II} \\
   & \hspace{-2em}
   =
   2 R_u R_v \, 
   \Bigl \{\, 
      m \cos(2\theta)
      + 
      \theta' 
      - 
      \frac{g^2}{\kappa+1}\, 
      [ R_u R_v ]^{\kappa+1} \,
   \Bigr \} \>.
   \notag   
\end{align}
In the next section, we give results for the s-s model and v-v model versions.

%
%
\subsection{\label{ss:ssModel}s-s model}

Using the results of the last section and after some algebra, we find equations for $R'_{u}$ and $R'_{v}$ to be given by:
\begin{subequations}\label{e:RRp-ss}
\begin{align}
   &R'_{u}
   +
   m R_{u} \sin(2 \theta)
   \label{e:Rp-ss-a} \\
   & \hspace{3em}
   -
   g^2 [\, R_u R_v \cos(2\theta) \,]^{\kappa} R_u \sin(2 \theta)
   = 
   0 \>,
   \notag \\
   &R'_{v}
   +
   m R_{v} \sin(2 \theta)
   \label{e:Rp-ss-b} \\
   & \hspace{3em}
   -
   g^2 [\, R_u R_v \cos(2\theta) \,]^{\kappa} R_v \sin(2 \theta)
   = 
   0 \>,
   \notag   
\end{align}
\end{subequations}
and an equation for $\theta'$ given by:
\begin{equation}\label{e:thetap}
   \theta'
   +
   m \cos(2 \theta)
   -
   \omega
   -
   g^2 [\, R_{u} \, R_{v} \,]^{\kappa} \cos^{\kappa+1}(2 \theta)
   =
   0 \>.
\end{equation}
Additionally, from \eqref{e:T11ss-conserved} we get
\begin{align}\label{e:T11ss-conserved-II}
   g^2 [\, R_{u} \, R_{v} \,]^{\kappa} \cos^{\kappa+1}(2 \theta)
   =
   (\kappa+1) \, [\, m \, \cos(2\theta) - \omega \,] \>.
\end{align}
Substitution of \eqref{e:T11ss-conserved-II} into \eqref{e:thetap} yields a simple differential equation for $\theta(x)$:
\begin{equation}\label{e:ss-thetaeq}
   \theta'
   =
   \kappa \, [ m \cos(2\theta) - \omega \,]\>.
\end{equation}
This equations is identical to Eq.~(22) in Ref.~\onlinecite{PhysRevE.82.036604}.  The solution is:
\begin{equation}\label{e:thetasol}
   \theta(x)
   =
   \tan^{-1}[\,\alpha \tanh(\kappa \beta x) \,]
\end{equation}
where
\begin{equation}\label{e:alphabeta}
   \alpha
   = 
   \sqrt{\frac{m - \omega}
              {m + \omega}} \qc
   \beta
   =
   \sqrt{m^2 - \omega^2} \>.
\end{equation}
We have 
\bq
m+\omega = \frac{\beta}{\alpha};~~ m-\omega = \alpha \beta.
\eq
A useful identity is:
\begin{equation}\label{e:cos2theta}
   \cos[2 \, \theta(x)]
   =
   \frac{1 - \alpha^2 \tanh^2(\kappa \beta x)}
        {1 + \alpha^2 \tanh^2(\kappa \beta x)} \>.
\end{equation}
We also have:
\bq
m \cos 2 \theta - \omega  = \frac {\beta_{k}^2} {k^2(\omega +m \cosh 2 \beta_{k} x)} \>.
\eq

Solving \eqref{e:T11ss-conserved-II} for the product $R_u R_v$ we find that the charge density $\rho(x)$ is given by:
\begin{align}\label{e:FindRuRv-II}
   &\rho(x) = R_{u}(x) \, R_{v}(x)
   \\
   & \hspace{1em}
   =
   \frac{1}{\cos(2\theta(x))}
   \Bigl [\,
      \frac{ (\kappa+1) [m \cos(2\theta(x)) - \omega] }
           { g^2 \cos(2 \theta(x)) }
   \Bigr ]^{1/\kappa} \>,
   \notag
\end{align}
Where $\cos(2\theta(x))$ is given by \eqref{e:cos2theta}.
Thus we can rewrite $\rho(x)$ in the form:
\bq
\rho(x)  = \frac  {1+\alpha^2 \tanh^2 \kappa \beta  x }  {1-\alpha^2 \tanh^2 \kappa \beta  x } 
  \left[ \frac{ (\kappa+1) (\kappa \beta)^2 \sech^2 \beta \kappa x  }{g^2 \kappa^2 (m+\omega) ( 1-\alpha^2 \tanh^2\beta \kappa x )} \right ]^{\frac{1}{\kappa}}  \>.
  \label{eq:rho}
\eq

Individual equations for $R_u(x)$ and $R_v(x)$ are obtained by substituting \eqref{e:T11ss-conserved-II} into \eqref{e:RRp-ss} and obtaining:
\begin{subequations}\label{e:RRp-ss-II}
\begin{align}
   R'_u - \kappa \, m \, R_u \sin(2\theta) 
   + 
   (\kappa+1) \, \omega R_n \tan(2\theta) 
   &= 0 \>,
   \label{e:RRp-ss-II-a} \\
   R'_v - \kappa \, m \, R_v \sin(2\theta) 
   + 
   (\kappa+1) \, \omega R_v \tan(2\theta) 
   &= 0 \>.
   \label{e:RRp-ss-II-b}   
\end{align}
\end{subequations}
Using the trig identities,
\begin{equation*}
   \sin(2\theta)
   =
   2 \tan(\theta) \cos^2(\theta)
   \qc
   \tan(2 \theta)
   =
   \frac{2 \tan(\theta)}{1 - \tan^2(\theta)} \>,   
\end{equation*}
and substitution into \eqref{e:RRp-ss-II} gives
\begin{align}
   \dv{\ln{R_u}}{x}
   &=
   \frac{2 \kappa \, m  \tan(\theta)}{1 + \tan^2(\theta)} 
   - 
   \frac{2 (\kappa+1) \, \omega \tan(\theta)}{1 - \tan^2(\theta)}
   \label{e:lnRu-ss} \\
   & \hspace{-2em}
   =
   \frac{2 \kappa \, m \, \alpha \tanh(\kappa \beta x)}
        {1 + \alpha^2 \tanh^2(\kappa \beta x)} 
   - 
   \frac{2 (\kappa+1) \, \omega \alpha \tanh(\kappa \beta x)}
        {1 - \alpha^2 \tanh^2(\kappa \beta x)} \>.
   \notag 
\end{align}
with an identical equation for $R_v$ with the substitution $R_u \rightarrow R_v$.  Integrating, we find:
\begin{align}
   \ln{\left(R_u\right)}
   &=
   \frac{m \alpha}{(\alpha^2 + 1) \beta}
   \label{e:Ru-ss-result} \\
   & \hspace{-1em}
   \times
   \ln[( 1 + \alpha^2 \tanh^2(\kappa \beta x)) \cosh^2(\kappa \beta x)]
   \notag \\
   & \hspace{-1em}
   -
   \frac{(\kappa+1) \omega \alpha}{\kappa (\alpha^2 - 1) \beta}
   \notag \\
   & \hspace{-1em}
   \times
   \ln[( 1 - \alpha^2 \tanh^2(\kappa \beta x)) \cosh^2(\kappa \beta x)]
   +
   C_u \>.
   \notag
\end{align}
where $C_u$ is an integration constant.  The solution for $\ln{\left(R_v\right)}$ is identical except for the integration constant, which becomes $C_v$.  So the functions $R_u(x)$ and $R_v(x)$ are the same except for an overall constant normalization factor.  The product $R_u(x) R_v(x)$ is a fixed function of $x$ given by  \eqref{e:FindRuRv-II}, which fixes the product of the constants of integration $C_uC_v$.

\begin{figure}[t]
   \centering
   \subfigure[\ charge density]
   { \label{f:fig1a}
     \includegraphics[width=0.9\linewidth]{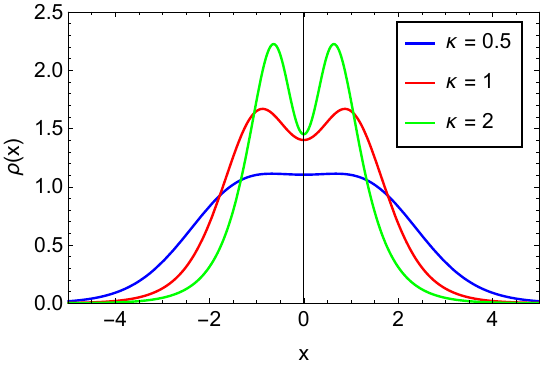} }
   \subfigure[\ energy density]
   { \label{f:fig1b}
     \includegraphics[width=0.9\linewidth]{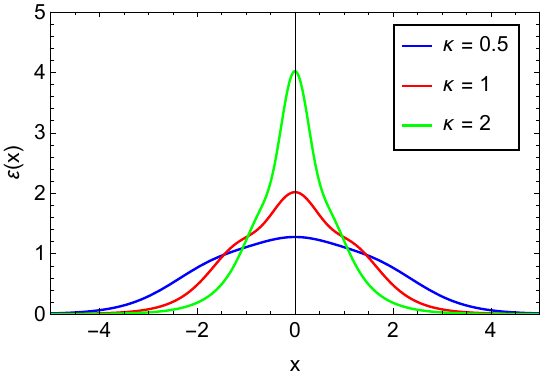} }
   \caption{\label{f:densities-ss} Plot of the charge density $\rho(x)$ (top) 
   and energy density $\varepsilon(x)$ (bottom) both as functions of $x$ for
   the s-s case with $g = 1$, and for the case when $m=1$ and 
   $\omega = 0.3$ for several values of $\kappa$.}
\end{figure}

From \eqref{e:T00ss-II}, the energy density for the s-s model is
\begin{align}
   \varepsilon_{\text{s-s}}(x)
   &=
   2 R_u R_v \, 
   \Bigl \{\, 
      (1 + \kappa) m \cos(2 \theta) - \kappa \omega  
      \label{e:varepsilon-ss} \\
      & \hspace{5em}  
            - 
      \frac{g^2 [ R_u R_v\cos(2 \theta) ]^{\kappa+1}}{\kappa+1}\, 
       \,
   \Bigr \} \>.
   \notag
\end{align}
Plots of the charge and energy densities as a function of $x$ for the s-s case with $m = 1$ and $\omega = 0.3$ are shown in Fig.~\ref{f:densities-ss} for several values of $\kappa$.
%
%
%
%


The total charge and energy can be found by integrating the densities over all $x$.
For the total charge, we find using the expression for $\rho(x) $ in Eq. \ref{eq:rho} and 
make the substitution:
\bq    \tanh( \kappa \beta x) \rightarrow y \eq
so that : 
\begin{equation}\label{C11-jd}
   Q
   =
   \int_{-\infty}^{\infty}\!\! \rho(x) \dd{x} 
   = 
   \frac{2}{\kappa \beta}
   \Bigl [\,
      \frac{ (\kappa+1)\beta^{2} }
           {g^2 (m +\omega)} 
   \Bigr ]^{1/\kappa}
   J_{\kappa}(\alpha^2) \>,
\end{equation}
where
\begin{align}
   J_{\kappa}(\alpha^2) 
   &=
   \int^{1}_{-1} dy \, \frac{1+\alpha^2 y^2}{(1-y^2)^{\frac{1}{\kappa}(\kappa-1)}(1-\alpha^2 y^2)^{\frac{1}{\kappa}(\kappa+1)}} \>. \notag \\
   &= 
   B\Bigl (\frac{1}{2},\frac{1}{\kappa} \Bigr ) \,
   _2F_1 
   \Bigl ( 
      \frac{\kappa+1}{\kappa},\frac{1}{2}, \frac{\kappa+2}{2\kappa};\alpha^2 
   \Bigr ) \label{12}\\ 
   & \qquad
   +
   \alpha^2 B\Bigl (\frac{3}{2} ,\frac{1}{\kappa} \Bigr ) \,
   _2F_1
   \Bigl (
      \frac{\kappa +1}{\kappa},\frac{1}{2},\frac{3\kappa+2}{2\kappa};\alpha^2
   \Bigr ) \>.
   \notag
\end{align}
To obtain the final result one makes the substitution $y \rightarrow t^{1/2} $ and uses the definition:

\bq
 \int_0^1  dt \ \frac{t^{b-1} (1-t) ^{c-b-1}}{(1-tz)^{a}}= \frac{\Gamma(b) \Gamma(c-b)}{\Gamma(c)} \ _2F_1(a,b,c;z) \>,
 \eq

In a similar fashion we find that the total energy is given by
\begin{equation}\label{e:E-ss}
   E
   =
   \int\!\! \varepsilon(x) \dd{x} 
   =
   E_1 + E_2 \>,
\end{equation}
where
\begin{subequations}\label{e:E1E2-ss}
\begin{align}
   E_1
   &=
   \frac{2\beta}{m+\omega} 
   \Bigl [ \,
      \frac{(\kappa+1)\beta_{k}^{2}}
      {g^2\kappa^2(m+\omega)}
   \Bigr ]^{1/\kappa} 
   B \Bigl ( \frac{1}{2},1+\frac{1}{\kappa} \Bigr )
   \label{5.21} \\ 
   & \qquad \times 
     {}_2 F_1 
     \Bigl ( 
        \frac{1+\kappa}{\kappa},\frac{1}{2},\frac{3\kappa+2}{2\kappa};-\alpha^2
     \Bigr ) \>,
   \notag \\
   E_2
   &=
   \frac{2}{\beta_k}[\frac{(\kappa+1)\beta_{k}^{2}}
      {g^2\kappa^2(m+\omega)}]^{1/\kappa} B(\frac{1}{2} ,\frac{1}{\kappa}) 
   \label{5.23} \\
   & \qquad \times
   {}_2F_1 
   \Bigl (
      \frac{1}{\kappa},\frac{1}{2},\frac{\kappa+2}{2\kappa};\alpha^2
   \Bigr ) \>.
   \notag
\end{align}
\end{subequations}
In units of $m$, the allowed region for $\omega$ is: $0 < \omega < 1$. 
The charge and energy are double that found in the single field case of Ref.~\onlinecite{PhysRevE.82.036604}, so that the results displayed there for $\omega/m$ and $H_{sol}/m$ at $Q=1$ as a function of $\kappa$ in Fig.~1 of Ref.~\onlinecite{PhysRevE.82.036604} applies to our results here.  The equation \eqref{C11-jd} for Q allows one to determine $\omega/m$ as a function of $g$ for various values of $\kappa$ at $m=1$.  We find in this case that the range of $g$ values for the existence of a bound state, as a function of $\kappa$, is bounded from below. The functional dependence of the lower bound $g_{\text{min}}$, together with the corresponding solution $\omega(g_{\text{min}})$, as a function of $\kappa$ are depicted in Fig.~2 of Ref.~\onlinecite{PhysRevE.82.036604}.  Summarizing, we find that in the s-s case, bound states exist for all values of $\kappa$ when $g > g_{\text{min}}$.

%
%
\subsection{\label{ss:vvModel}v-v model}

The corresponding equations for $R'_u$ and $R'_v$ for the v-v model are given by:
\begin{subequations}\label{e:additvv}
\begin{align}
   R'_{u}
   +
   m R_{v} \sin(2 \theta)
   &= 
   0 \>,
   \label{e:additvv-a} \\
   R'_{v}
   +
   m R_{u} \sin(2 \theta)
   &= 
   0 \>,
   \label{e:additvv-b}   
\end{align}
\end{subequations}
and the equation for $\theta'$ is given by
\begin{equation}\label{e:subitvv}
   \theta' + m \cos(2 \theta) - \omega - g^2 [\, R_u R_v ]^{\kappa} = 0 \>.
\end{equation}
From \eqref{e:T11vv-II}, we get:
\begin{equation}\label{e:T11vv-III}
   g^2 [R_u R_v]^{\kappa}
   =
   (\kappa + 1) \, [\, m \cos(2 \theta) - \omega ]\,] \>.
\end{equation}
Substitution of \eqref{e:T11vv-III} into \eqref{e:subitvv} gives
\begin{equation}\label{e:vv-thetaeq}
   \theta'
   =
   \kappa \, [ m \cos(2\theta) - \omega \,]\>,
\end{equation}
which is the same equation we found in \eqref{e:ss-thetaeq} for the s-s coupling case and has the same solution given in \eqref{e:thetasol}.  The charge density for the v-v case is given here by
\begin{align}
   \rho(x)
   &= 
   R_u(x) R_v(x)
   \label{e:RuRv-vv} \\
   &=
   \{\, (\kappa + 1) \, [\, m \cos(2 \theta(x)) - \omega ] / g^2 \, \}^{1/\kappa}
   \notag \\
   &=
   \Bigl [\,
      \frac{(\kappa+1) \, \alpha \beta \, \sech^2(\kappa \beta x)}
      { g^2 [1 + \alpha^2 \tanh^2(\kappa \beta x)] } \,      
   \Bigr ]^{1/\kappa}
   \notag
\end{align}
where we have used \eqref{e:cos2theta}.  To find $R_u(x)$ and $R_v(x)$, we rewrite  Eqs.~\eqref{e:additvv} using the trig identities to get:
\begin{subequations}\label{e:additvv-sol}
\begin{align}
   \dv{\ln{R_u}}{x}
   &=
   -
   \frac{ 2 m \alpha \tanh(\kappa \beta x)}
        {1 + \alpha^2 \tanh^2(\kappa \beta x)} \>,
   \label{e:additvv-sol-a} \\
   \dv{\ln{R_v}}{x}
   &=
   -
   \frac{ 2 m \alpha \tanh(\kappa \beta x)}
        {1 + \alpha^2 \tanh^2(\kappa \beta x)} \>,
   \label{e:additvv-sol-b}   
\end{align}
\end{subequations}
These equations have solutions given by:
\begin{subequations}\label{e.2:vv-RuRv-II}
\begin{align}
   R_u(x)
   &=
   C_u \, 
   [\, \cosh^2(\kappa \beta x) 
       + \alpha^2 \sinh^2(\kappa \beta x) \,
   ]^{1/(2\kappa)} \>,
   \\
   R_v(x)
   &=
   C_v \, 
   [\, \cosh^2(\kappa \beta x) 
       + \alpha^2 \sinh^2(\kappa \beta x) \,
   ]^{1/(2\kappa)} \>,
\end{align}
\end{subequations}
where $C_u$ and $C_v$ are constants of integration.  These constants are then fixed by the product relation \eqref{e:RuRv-vv}.  So we conclude that the two soliton fields $\Psi(x)$ and $\Phi(x)$ are proportional to each other.

\begin{figure}[t]
   \centering
   \subfigure[\ charge density]
   { \label{f:fig2a}
     \includegraphics[width=0.9\linewidth]{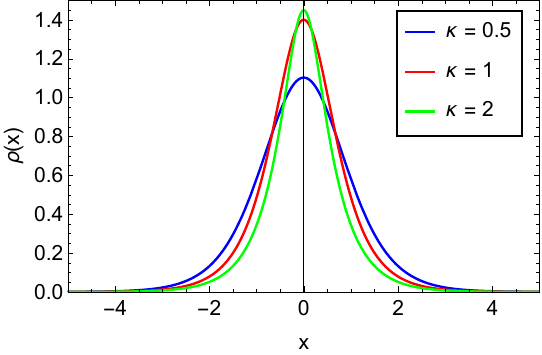} }
   \subfigure[\ energy density]
   { \label{f:fig2b}
     \includegraphics[width=0.9\linewidth]{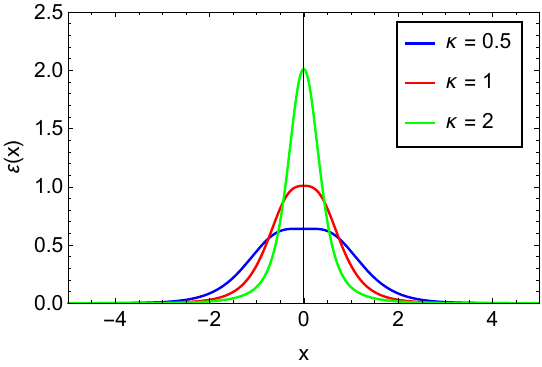} }
   \caption{\label{f:densities-vv} Plot of the charge density $\rho(x)$ (top) 
   and energy density $\varepsilon(x)$ (bottom) both as functions of $x$ for
   the v-v case with $g = 1$, and for the case when $m=1$ and 
   $\omega = 0.3$ for several values of $\kappa$.}
\end{figure}

For the v-v model, from \eqref{e:T00vv-II}, the energy density is given by:
\begin{align}
   \varepsilon_{\text{v-v}}(x)
   &=
   2 R_u R_v \, 
   \Bigl \{\, 
      (1 + \kappa) m \cos(2\theta)
      - 
      \kappa \omega
      \label{e:varepsilon-vv} \\
      & \hspace{5em} 
      - 
      \frac{g^2}{\kappa+1}\, 
      [ R_u R_v ]^{\kappa+1} \,
   \Bigr \} \>.  
   \notag 
\end{align}   
Plots of the charge and energy densities as a function of $x$ for the v-v case with $m = 1$ and $\omega = 0.3$ are shown in Fig.~\ref{f:densities-vv} for several values of $\kappa$.

%
%
%
%

For the v-v case, the total charge and energy is again obtained by integrating over the space dimension.  For the charge, we find:
\begin{equation}\label{C4-jd}
   Q
   =
   \int_{-\infty}^{\infty}\!\! \rho(x) \dd{x} 
   =
   \frac{2}{\kappa \beta}
   \Bigl [
      \frac{(\kappa+1)\beta^{2}}
           {g^2 (m+\omega)}
   \Bigr ]^{1/\kappa} I_{\kappa}(\alpha^2) \>,
\end{equation}
where
\begin{equation}\label{C5}
   I_{\kappa}(\alpha^2) 
   = 
   B \Bigl ( \frac{1}{2},\frac{1}{\kappa} \Bigr ) \,
   {}_2 F_1 
   \Bigl (
      \frac{1}{2},\frac{1}{\kappa},\frac{\kappa+2}{2\kappa};-\alpha^2
   \Bigr ) \>,
\end{equation}
and for the energy, we get:
\begin{equation}\label{e:E-vv}
   E
   =
   \int\!\! \varepsilon(x) \dd{x} 
   =
   (1 - 1/\kappa) E_1 + E_2 \>,
\end{equation}
where
\begin{subequations}\label{e:E1E2-vv}
\begin{align}
   E_1
   &= 
   \frac{2\beta}{(m+\omega)} 
   \Bigl [
      \frac{(\kappa+1)\beta^{2}}
        {g^2 (m+\omega)}
   \Bigr ]^{1/\kappa} \,  
   B \Bigl (\frac{1}{2} ,\frac{1+\kappa}{\kappa} \Bigr )  
   \label{C9-jd} \\
   & \qquad \times 
   {}_2F_1 
   \Bigl (
      \frac{1+\kappa}{\kappa},\frac{1}{2} ,\frac{3\kappa+2}{2\kappa};-\alpha^2
   \Bigr ) \>,
   \notag \\
   E_2
   &=
   \frac{4}{\beta_k}
   \Bigl [ \frac{(\kappa+1)\beta_{k}^{2}}
                {g^2\kappa^2(m+\omega)}
   \Bigr ]^{1/\kappa} B \Bigl( \frac{1}{2},\frac{1}{\kappa} \Bigr )
   \label{C9a-jd} \\ 
   & \qquad \times 
   \Bigl \{
      {}_2F_1 
      \Bigl (
         \frac{1}{\kappa},\frac{1}{2},\frac{\kappa+2}{2\kappa};-\alpha^2 
      \Bigr )
   \notag \\
   & \qquad\qquad  
   -
   {}_2F_1 
   \Bigl (
      \frac{1}{2},\frac{1}{\kappa},\frac{\kappa+2}{2\kappa};-\alpha^2
   \Bigr ) 
   \Bigr \} \>.
   \notag
\end{align}
\end{subequations}
Here $0 < \omega < m$, and $E$ is in units of mass, so we can just set $m=1$ in the following.  
Note while $E$ and $Q$ are twice the value of the single component case, the ratio $E/Q$ remains unchanged and hence Figs.~3 and 4 of Ref.~\onlinecite{PhysRevE.82.036604} are also figures for the present problem once we set $Q=1$ for our problem.  When we set $Q=1$ we can solve for $\omega/m$ in terms of $g$ and $\kappa$  from Eq.~\eqref{C4-jd} for Q.  As in the single field case, in order for a bound state to 
exist $E_{sol} < m$, or $E/m <1 $.  This puts constraints on the allowed values of $g$, so there is both a minimum and maximum value of $g$ as a function of $\kappa$.  Fig.~3 of Ref.~\onlinecite{PhysRevE.82.036604} maps out the allowed values of $\omega$ and $g^2$ for various values of $\kappa$ for $M=1$. The allowed range of $g$ values for the existence of a bound state, as a function of $\kappa$, has both a lower and an upper bound, and the domain shrinks as $\kappa$ increases.  Around $\kappa=2.5$, these bounds cross, and no bound states are possible for $ \kappa > 2.5$. The functional dependence of $g_{\text{min}}$ and $g_{\text{max}}$, together with the corresponding solutions $\omega(g_{min})$ and $\omega(g_{max})$, as a function of $\kappa$, are depicted in Fig.~4 of \cite{PhysRevE.82.036604}.
 
%
%
\section{\label{s:alternatives}Alternative solutions}

We have tried other parameterizations for the wave functions $\Phi,\Psi$, but because $j_1(x)=0$ and $T_{11}(x) =0$, the final expressions for the product $R_uR_v$ and $g(x) = \tan \theta(x)$ are unchanged. 

%
%
\section{\label{s:NonrelReduction}Nonrelativistic limit}

The full component equations are given by Eq.~\eqref{e:ss-eom-II} for the s-s case and by Eq.~\eqref{e:vv-eom-II} for the v-v case.  One can write the coupled NLD equations (NLDEs) in the s-s case as
\begin{subequations}\label{e:EOM-ss}
\begin{align}
    \rm i \sigma_3 \partial_t \Psi + \sigma_1 \partial_1 \Psi - m \Psi - V_I \Psi &= 0 \>,
    \label{e:EOM-ss-a} \\
    \rm i \sigma_3 \partial_t \Phi + \sigma_x \partial_1 \Phi - m \Phi - V_I \Phi &= 0 \>,
    \label{e:EOM-ss-b}
\end{align}
\end{subequations}
where $V_I = g^2(\bar{\Phi} \Psi)^\kappa$.  We see that for the s-s case, $\Phi$ and $\Psi$ obey the same equations, thus we come to the same conclusion the the fields $\Phi$ and $\Psi$ must be proportion. 

 For the v-v case we get a more complicated structure.
One can write the coupled NLDE equations as 
\begin{subequations}\label{e:EOM-vv}
\begin{align}
   \rm i \sigma_3 \partial_t \Psi + \sigma_x \partial_x \Psi - m \Psi - V_I[ \Psi,\Phi] \Phi
   &= 0 \>,
   \label{e:EOM-vv-a} \\
   \rm i \sigma_3 \partial_t \Phi + \sigma_x \partial_x \Phi - m \Phi - V_I [\Phi ,\Psi] \Psi
   &= 0 \>,
   \label{e:EOM-vv-b}
\end{align}
\end{subequations}
where
\begin{subequations}\label{e:VI-vv}
\begin{align}
   V_I[ \Psi,\Phi] 
   &=
   g^2 (u_1^2 + u_2^2 )^{(\kappa+1)/2} (v_1^2 + v_2^2 )^{(\kappa-1)/2} \>,
   \label{e:VI-vv-a} \\
   V_I[ \Phi,\Psi] 
   &= 
   g^2 (v_1^2 + v_2^2 )^{(\kappa+1)/2} (u_1^2 + u_2^2 )^{(\kappa-1)/2} \>.
   \label{e:VI-vv-b}
\end{align}
\end{subequations}

In the NR reduction as suggested by Moore \cite{SCOTT1990366} one writes the nonlinear interaction 
term $V_{I}$ as
\be\label{6}
   V_{I}(\lambda) = \frac{(1+\sigma_3)}{2} V_{I} 
   +\lambda \frac{(1-\sigma_3)}{2} V_{I}\,,
\ee
and then does perturbation theory in $\lambda$. 

If $(u_0,v_0)^{T}$ and $(\phi_{10},\phi_{20})$ corresponds to solution at $\lambda = 0$ then
the above four equation in the lowest approximation reduce to
\begin{subequations}\label{e:NRreduction}
\begin{align}
   &\dv{u_0}{x} + (m+\omega) v_0 
   = 0 \>,
   \label{e:NRreduction-a} \\
   & 
   \dv{v_0}{x} + (m - \omega) u_0 
   \label{e:NRreduction-b} \\
   & \hspace{-1em}
   - 
   g^2 
   ( v_{0}^{2} + u_{0}^{2})^{(\kappa+1)/2}
   (\phi_{10}^{2}+\phi_{20}^{2})^{(\kappa-1)/2} \phi_{10} 
   = 0 \>,
   \notag \\
   &\dv{\phi_{10}}{x} + (m+\omega) \phi_{20} 
   = 0 \>,
   \label{e:NRreduction-c} \\
   &\frac{d\phi_{20}}{dx} + (m-\omega) \phi_{10} 
   \label{e:NRreduction-d} \\
   & \hspace{-1em} 
   - 
   (v_{0}^{2} +u_{0}^{2})^{(\kappa-1)/2} 
   (\phi_{10}^{2}+\phi_{20}^{2})^{(\kappa+1)/2}u_0 
   = 0 \>.
   \notag
\end{align}
\end{subequations}
On differentiating \eqref{e:NRreduction-a} and using \eqref{e:NRreduction-b}, we obtain
\begin{equation}\label{e:unusual-1}
   - \frac{1}{2m} \, \dv[2]{u_0}{x}
   +
   \hat{V_{1}}(x) \, \phi_{10}
   =
   \hat{E} \, u_0 \>.
\end{equation}

\be\label{11}
   -\frac{u_{0xx}}{2m} + \hat{V_{1}} \phi_{10} = \hat{E} u_0 \>,
\ee
where
\be\label{12_2}
   \hat{E} = \epsilon_0 \Bigl ( 1 + \frac{\epsilon_0}{2m} \Bigr )
   \qc
   \epsilon_0 = \frac{\omega-m}{2m} \>,
\ee
and 
\be\label{13}
   \hat{V_{1,2}} = V_{1,2} \Bigl ( 1+\frac{\epsilon_0}{2m} \Bigr ) \>,
\ee
with
\begin{subequations}\label{e:V12}
\begin{align}
   V_1 
   &= g^2 
      (u_{0}^{2}+v_{0}^{2})^{(\kappa+1)/2} 
      (\phi_{10}^{2}+\phi_{20}^{2})^{(\kappa-1)/2} \>,
   \label{e:V1} \\
   V_2
   &= g^2
       (u_{0}^{2}+v_{0}^{2})^{(\kappa-1)/2}
       (\phi_{10}^{2}+\phi_{20}^{2})^{(\kappa+1)/2} \>.
   \label{e:V2}
\end{align}
\end{subequations}
Similarly, on differentiating \eqref{e:NRreduction-c} and using \eqref{e:NRreduction-d} we obtain
\begin{equation}\label{e:unusual-2}
   - \frac{1}{2m} \, \dv[2]{\phi_{10}}{x}
   +
   \hat{V_{2}}(x) \, u_0
   =
   \hat{E} \, \phi_{10} \>.
\end{equation}
Thus we have obtained two unusual NLS as given by \eqref{e:unusual-1} and \eqref{e:unusual-2}.
These coupled equations are highly unusual since
\begin{enumerate}
\item{While the fields $v_0$ and $\phi_{20}$ appear in \eqref{e:unusual-1} and 
\eqref{e:unusual-2} respectively, these fields are not dynamical fields in the sense that
their second derivatives do not appear anywhere.}
\item{While in \eqref{e:unusual-2} with second derivative term $u_{0xx}$, 
the $\phi_{10}$ field appears multiplied by the nonlinear term, whereas the term
$u_0$ appears in \eqref{e:unusual-2} with second derivative term 
$\phi_{10xx}$ term. Again in the nonrelativistic reduction for the v-v case one finds that  $\phi_{10}
\propto u_{0}$.}
\end{enumerate}

%
%
\section{\label{s:Conclusion}Conclusions}

Inspired by the coupled massive Thirring model introduced recently by 
Basu-Mallick et al \cite{BasuMallick2023bosonic}, in this paper we have 
introduced a coupled Soler model.  Further, we have generalize both the coupled 
MTM and the coupled Soler model to arbitrary nonlineariy parameter $\kappa$ and 
found exact solutions in both the cases.  Remarkably, as a result of the 
conservation laws inherent in the the Lagrangian, the exact solutions for the 
two coupled fields turn out to have the two fields proportional to one another.
Thus the solutions we found in both the models are related to those that we had 
already found for the uncoupled (single field) models in Ref.~\onlinecite{PhysRevE.82.036604}. 

There are few open questions. The most important being about the stability of
the solutions that we have obtained in both the coupled models. In this context
it is worth recalling that the stability of the solutions of the two uncoupled 
(single field) models was discussed in Ref.~\onlinecite{PhysRevE.90.036604} where it was 
found that all the two humped solutions were unstable while the stability of the
single humped solutions depended on the value of $\kappa$.  In that paper, the 
stability regimes were found by direct simulation of the NLDE. However, it will
be worth exploring the spectrum of the associated linearized operator in the
spirit of~\cite{COMECH2014639}.~This imposes its own challenges from the numerical
analysis perspective, as suitable numerical discretizations schemes should be used
for handling first-order derivative operators that will appear in the linearization.%
~The other question is about the integrability of the uncoupled as well as coupled
Soler model in case $\kappa = 1$. Recall that the corresponding coupled as 
well as the uncoupled MTM are integrable.  So far as we are aware of, the 
uncoupled Soler model is not integrable. We believe that these questions
are worth pursuing. We hope to address some of these issues in future.

%
%
\acknowledgments

One of us (AK) is grateful to Indian National Science Academy (INSA) for the 
award of INSA Honorary Scientist position at Savitribai Phule Pune University. 
FC, JFD, and EGC would like to thank the Santa Fe Institute and the Center for 
Nonlinear Studies at Los Alamos National Laboratory for their kind hospitality.
The work at Los Alamos National Laboratory was carried out under the auspices of
the U.S.~DOE and NNSA under Contract No.~DEAC52-06NA25396.

%
%
\appendix
%
%
\section{\label{s:StressEnergyTensor}Lagrangian and stress-energy tensor}

The stress-energy tensor \eqref{e:Tdef} is defined by:
\begin{align}\label{a:Tdef}
   T^{\mu\nu} &= \frac{1}{2} \, \{\, D^{\mu\nu} + \text{h.c.} \,\} - g^{\mu\nu} L \>,
   \\
   D^{\mu\nu} 
   &=
   \rmi \, \barPhi \gamma^{\mu} \partial^{\nu} \Psi
   +
   \rmi \, \barPsi \gamma^{\mu} \partial^{\nu} \Phi\>,
\end{align}
where $L$ is given by \eqref{e:Lagrangian}. 
We can write the equations of motion  as
\begin{equation} \label{fred1}
\frac{\delta L} {\delta \barPhi} =   (\rmi \gamma^{\mu} \partial_{\mu} - m) \Psi 
+ \frac{\delta L_I} {\delta \barPhi}=0,
\end{equation} 
\begin{equation} \label{fred2} 
\frac{\delta L} {\delta \barPsi} =   (\rmi \gamma^{\mu} \partial_{\mu} - m) \Phi 
+ \frac{\delta L_I} {\delta \barPsi}=0.
\end{equation} 
Multiplying  Eq. \eqref{fred1} by $\barPhi$ and Eq. \eqref{fred2} by $\barPsi$ and adding, we get the useful identity valid for both s-s and v-v interactions:
\begin{equation} \label{useful} 
    \barPhi (\rmi \gamma^{\mu} \partial_{\mu} - m) \Psi 
    + 
    \barPsi (\rmi \gamma^{\mu} \partial_{\mu} - m) \Phi + (\kappa+1) L_I
    = 0 \>.
\end{equation} 
The Hamiltonian density is given by
\begin{align}
   \calH
   &=
   T_{00}
   =
   h_1 + h_2 - h_3 
   \label{e:calH} \\
   &=
   \bar \Phi (\rmi \gamma_1 \partial_1) \Psi 
   + 
   \bar \Psi ( \rmi \gamma_1 \partial_1) \Phi 
   + 
   m (\bar \Psi \Phi+ \bar \Phi \Psi) 
   - 
   L_I.
   \notag
\end{align}
But since $h_1 = \kappa L_I = \kappa h_3$, we find that 
\bq
   \calH = h_2 + (\kappa-1) h_3= h_1(1-1/\kappa) + h_2 \>.
\eq

For the solitary wave ansatz we obtain for $u_i,v_i$ the results
\begin{align*}
   \barPhi (\rmi \gamma^{\mu} \partial_{\mu} - m) \Psi
   &=
   \omega \, (v_1 u_1 + v_2 u_2) 
   \\
   &\hspace{-2em} 
   - 
   ( v_1 u'_2 - v_2 u'_1 ) 
   - 
   m \, (v_1 u_1 - v_2 u_2 ) \>,
   \notag \\
   \barPsi (\rmi \gamma^{\mu} \partial_{\mu} - m) \Phi
   &=
   \omega \, (u_1 v_1 + u_2 v_2) 
   \\
   &\hspace{-2em} 
   - 
   ( u_1 v'_2 - u_2 v'_1 ) 
   - 
   m \, (u_1 v_1 - u_2 v_2 ) \>.
   \notag
\end{align*}
The self-interaction terms $L_I$ work out to be
\begin{align*}
   L_I^{\text{s-s}}
   &=
   \frac{2 g^2}{\kappa+1}\, (v_1 u_1 - v_2 u_2)^{\kappa+1} \>,
   \\
   L_I^{\text{v-v}}
   &=
   \frac{2 g^2}{\kappa+1}
   \bigl [\,
      ( v_1^2 + v_2^2 )( u_1^2 + u_2^2 )
   \bigr ]^{(\kappa+1)/2} \>.   
\end{align*}
Combining these two results, the Lagrangians for the s-s and v-v models are given by:
\begin{subequations}\label{a:LssLvv}
\begin{align}
   L_{\text{s-s}}
   &=
   2 \omega \, (u_1 v_1 + u_2 v_2)
   -
   2 m \, (u_1 v_1 - u_2 v_2)
   \label{a:Lss} \\
   & \hspace{1em}
   +
   u'_1 v_2 - u'_2 v_1 + v'_1 u_2 - v'_2 u_1
   \notag \\
   & \hspace{2em}
   +
   \frac{2 g^2}{\kappa+1}\, (v_1 u_1 - v_2 u_2)^{\kappa+1} \>,
   \notag \\
   L_{\text{s-s}}
   &=
   2 \omega \, (u_1 v_1 + u_2 v_2)
   -
   2 m \, (u_1 v_1 - u_2 v_2)
   \label{a:Lvv} \\
   & \hspace{-1em}
   +
   u'_1 v_2 - u'_2 v_1 + v'_1 u_2 - v'_2 u_1
   \notag \\
   & \hspace{-1em}
   +
   \frac{2 g^2}{\kappa+1}
   \bigl [\,
      ( v_1^2 + v_2^2 )( u_1^2 + u_2^2 )
   \bigr ]^{(\kappa+1)/2} \>.
   \notag 
\end{align}
\end{subequations}

For the stress-energy tensor, we will need the following results:
\begin{align*}
   D^{00}
   &=
   2 \omega \,( \, v_1 u_1 + v_2 u_2 \,)
   \\
   D^{11}
   &=
   v_1 u'_2 - v_2 u'_1 + u_1 v'_2 - u_2 v'_1 \>,
   \\
   D^{01}
   &=
   - \rmi \,[\, v_1 u'_1 + v_2 u'_2 + u_1 v'_1 + u_2 v'_2 \,] \>,
   \\
   D^{10}
   &= 0
   \notag
\end{align*}
Adding the complex conjugate, we find that $T^{01}=T^{10}=0$ for both the s-s and v-v models.  For the s-s model,
\begin{align}
   T^{00}_{\text{s-s}}
   &=
   2 m \, (u_1 v_1 - u_2 v_2)
   -
   (u'_1 v_2 - u'_2 v_1) 
   \label{e:T00-ss} \\
   & \hspace{-1em}
   - 
   (v'_1 u_2 - v'_2 u_1)
   -
   \frac{2 g^2}{\kappa+1}\, (v_1 u_1 - v_2 u_2)^{\kappa+1} \>,
   \notag \\
   T^{11}_{\text{s-s}}
   &=
   2 \omega \, (u_1 v_1 + u_2 v_2)
   -
   2 m \, (u_1 v_1 - u_2 v_2)
   \label{e:T11-ss} \\
   & \hspace{3em}
   +
   \frac{2 g^2}{\kappa+1}\, (v_1 u_1 - v_2 u_2)^{\kappa+1} \>,
   \notag
\end{align}
whereas for the v-v model, we have
\begin{align}
   T^{00}_{\text{v-v}}
   &=
   2 m \, (u_1 v_1 - u_2 v_2)
   -
   (u'_1 v_2 - u'_2 v_1) 
   \label{e:T00-vv} \\
   & \hspace{-1em}
   - 
   (v'_1 u_2 - v'_2 u_1)
   -
   \frac{2 g^2}{\kappa+1}
   \bigl [\,
      ( v_1^2 + v_2^2 )( u_1^2 + u_2^2 )
   \bigr ]^{(\kappa+1)/2} \>,
   \notag \\
   T^{11}_{\text{v-v}}
   &=
   2 \omega \, (u_1 v_1 + u_2 v_2)
   -
   2 m \, (u_1 v_1 - u_2 v_2)
   \label{e:T11-vv} \\
   & \hspace{1em}
   +
   \frac{2 g^2}{\kappa+1}
   \bigl [\,
      ( v_1^2 + v_2^2 )( u_1^2 + u_2^2 )
   \bigr ]^{(\kappa+1)/2} \>.
   \notag 
\end{align}

%
\bibliography{MTM.bib}
%
%
\end{document}

%% file: JohnsREVTEXdefs.tex
\usepackage{graphicx}          
\usepackage{subfigure}         
\usepackage{dcolumn}           
\usepackage{capt-of}           
\usepackage{makecell}          
\usepackage{bbm}               
\usepackage{bm}                
\usepackage{time}              
\usepackage[mathlines]{lineno} 
\usepackage{stackrel}          
\usepackage{physics}           
\usepackage{amsmath}
\usepackage[mathscr]{eucal}    
\usepackage{hyperref}          
\hypersetup{
    pdfnewwindow=true,         
    colorlinks=true,           
    linkcolor=red,             
    citecolor=blue,            
    filecolor=green,           
    urlcolor=cyan              
}
\newcommand{\rmi}{{\rm i}}                     
\newcommand{\rme}{{\rm e}}                     
\newcommand{\Partial}[4]
   {\Bigl ( \frac{\partial #1 }{\partial #2 } \Bigr )_{\! #3, #4 }}
\newcommand{\calH}{\mathcal{H}}

\newcommand{\Tproduct}[1]%
   {\ensuremath{\mathcal{T} \{ \, #1 \, \} } }